\documentstyle[fleqn,epsf,graphicx]{article}
\begin{document}
\title{On the Phase Diagram of Spin-Polarized Attractive Hubbard Model: Weak Coupling Limit}
\author{A. Kujawa and R. Micnas}
\date{}

\maketitle

\vspace{-1.1cm}
\begin{center}
\begin{footnotesize}
\emph{Solid State Theory Division, Faculty of Physics,\\ Adam Mickiewicz University,\\ Umultowska 85,
61-614 Pozna\'n, Poland}
\end{footnotesize}
\end{center}

\begin{center}
\begin{footnotesize}
PACS numbers: 71.10.Fd, 74.20.Rp, 71.27.+a, 71.10.Hf
\end{footnotesize}
\end{center}

\begin{abstract}
\noindent The superfluid properties of the attractive Hubbard model in a Zeeman magnetic field, and in the weak coupling regime have been investigated. The temperature and magnetic field dependencies of the order parameter have been analyzed. Furthermore, the temperature vs. magnetic field and temperature vs. spin polarization phase diagrams for the 2D and 3D lattices have been obtained. For some parameters a reentrant transition has been found.  
\end{abstract}

\section{Introduction}
Spin-polarized superfluidity (in the context of the cold atomic Fermi gases) and unconventional superconductivity with a nontrivial Cooper pairing has recently been investigated both theoretically and experimentally \cite{matsuda}.

Due to the presence of a magnetic field, the numbers of particles with spin down and spin up are different. This makes the formation of Cooper pairs across the spin-split Fermi surface with non-zero total momentum ($\vec{k} \uparrow$, $-\vec{k}+\vec{q} \downarrow$) (Fulde and Ferrell \cite{fulde}, and Larkin and Ovchinnikov \cite{larkin} (FFLO) state) possible. The pairing in spin-polarized state is very interesting not only in the context of superconductivity, but also in that of trapped unbalanced ultracold Fermi atomic gases and color superconductivity in high energy physics \cite{gruenberg}-\cite{Casalbuoni}. 

In this paper we analyze the influence of a pure Zeeman effect on the superfluid characteristics within a lattice fermion (the spin-polarized attractive Hubbard) model. For sufficiently high magnetic fields we can observe a change in the character of finite temperature transition between the superconducting (SC) and the normal state (NS), from the second to the first order. Such a phenomenon has been discovered for the first time by Sarma in 1963 \cite{sarma}. This paper consists of three parts. The first part gives analysis of the spin-polarized attractive Hubbard model in the Hartree-Fock approximation. The second part presents numerical results. In sec. III we summarize the discussion. 

\section{Model}
The model Hamiltonian is the attractive Hubbard model ($U<0$) in a magnetic field \cite{MicnasModern}:
\begin{equation}
\label{ham}
H=\sum_{ij\sigma} (t_{ij}^{\sigma}-\mu \delta_{ij})c_{i\sigma}^{\dag}c_{j\sigma}+U\sum_{i} n_{i\uparrow}n_{i\downarrow}-h\sum_{i}(n_{i\uparrow}-n_{i\downarrow}),
\end{equation}
where: $n_{i\uparrow}=c_{i\uparrow}^{\dag}c_{i\uparrow}$, $n_{i\downarrow}=c_{i\downarrow}^{\dag}c_{i\downarrow}$, $t_{ij}^{\sigma}$ -- hopping integral, $U$ -- the on-site interaction, $\mu$ -- the chemical potential. The Zeeman field $h$ can be created by an external magnetic field (in ($g \mu_B \slash 2$) units) or by a population imbalance in the context of the cold atomic Fermi gases. 

Transforming the Hamiltonian (\ref{ham}) to the reciprocal space, one obtains:
\begin{eqnarray}
\label{ham'}
H&=&\sum_{\vec{k},\sigma}(\epsilon_{\vec{k}}^{\sigma}-\mu)c_{\vec{k}\sigma}^{\dag}c_{\vec{k}\sigma}+\frac{U}{N}\sum_{\vec{k_1},\vec{k_2},\vec{q}}c_{\vec{k_1}\uparrow}^{\dag}c_{\vec{k_1}-\vec{q}\uparrow}c_{\vec{k_2}\downarrow}^{\dag}c_{\vec{k_2}+\vec{q}\downarrow}-\nonumber\\
&-&h\sum_{\vec{k}} (c_{\vec{k}\uparrow}^{\dag}c_{\vec{k}\uparrow}-c_{\vec{k}\downarrow}^{\dag}c_{\vec{k}\downarrow}),
\end{eqnarray}
where the electron dispersion (with hopping only between the nearest neighbors) is $\epsilon_{\vec{k}}^{\sigma}=\sum_{\vec{\delta}}t_{\vec{\delta}}^{\sigma}e^{\vec{k} \cdot \vec{\delta}}=-2t^{\sigma} \Theta_{\vec{k}}$, $\Theta_{\vec{k}}=\sum_{l=1...d} cos(k_l a_l)$ (here $d=2,3$ for two- and three-dimensional lattice, respectively), $a_l$ is the lattice constant in the $l$-th direction (we set $a_l=1$ in further considerations).
Now, \mbox{using} the Hartree-Fock approximation (the pairing only with $\vec{q}=0$), we can obtain the following equation for the superconducting order parameter ($\Delta=-\frac{U}{N}\sum_{\vec{k}} \langle c_{-\vec{k} \downarrow} c_{\vec{k} \uparrow} \rangle$):
\begin{equation}
\label{del}
 \Delta=-\frac{U}{N}\sum_{\vec{k}}\frac{\Delta}{2\omega_{\vec{k}}}\frac{1}{2}\Bigg(\tanh\frac{\beta E_{\vec{k}\uparrow}}{2}+\tanh \frac{\beta E_{\vec{k}\downarrow}}{2}\Bigg),
\end{equation}
where:
\begin{equation}
E_{\vec{k}\downarrow}= (-t^{\downarrow}+t^{\uparrow})\Theta_{\vec{k}}+ \frac{UM}{2}+h+\omega_{\vec{k}},
\end{equation}
\begin{equation}
E_{\vec{k}\uparrow}=(-t^{\uparrow}+t^{\downarrow})\Theta_{\vec{k}}-\frac{UM}{2}-h+\omega_{\vec{k}},
\end{equation}
\begin{equation}
\omega_{\vec{k}}=\sqrt{((-t^{\uparrow}-t^{\downarrow})\Theta_{\vec{k}}-\bar{\mu})^2+|\Delta|^2},\qquad  \bar{\mu}=\mu-\frac{Un}{2},
\end{equation}
$\beta=1/k_B T$, $M=n_{\uparrow}-n_{\downarrow}$ -- spin magnetization (polarization), $n_{\sigma}=\frac{1}{N} \sum_{\vec{k}} \langle c_{\vec{k} \sigma}^{\dag} c_{\vec{k} \sigma} \rangle$, $n=n_{\uparrow}+n_{\downarrow}$ -- electron concentration.

Equation (\ref{del}) takes into account the spin polarization in the presence of a magnetic field and the spin-dependent hopping ($t^{\uparrow}\neq t^{\downarrow}$) \cite{Wilczek}.

The particle number equation takes the form:
\begin{equation}
n=1-\frac{1}{2N}\sum_{\vec{k}} \frac{-(t^{\uparrow}+t^{\downarrow})\Theta_{\vec{k}}-\bar{\mu}}{\omega_{\vec{k}}}\Bigg(\tanh\frac{\beta E_{\vec{k}\uparrow}}{2}+\tanh \frac{\beta E_{\vec{k}\downarrow}}{2}\Bigg).
\end{equation}
The equation for the magnetization is:
\begin{equation}
 M=\frac{1}{2N}\sum_{\vec{k}}\Bigg(\tanh\frac{\beta E_{\vec{k}\downarrow}}{2}-\tanh \frac{\beta E_{\vec{k}\uparrow}}{2}\Bigg).
\end{equation}
By calculating the partition function in the usual way, we can determine the grand canonical potential:
\begin{eqnarray}
\label{pot}
&&\hspace{-0.5cm}\frac{\Omega}{N}=\frac{1}{4} Un(2-n)-\mu+\frac{1}{4}UM^2 -\frac{|\Delta|^2}{U}\\
&&\hspace{-0.5cm}-\frac{1}{\beta N}\sum_{\vec{k}} \ln \Bigg (2\cosh\frac{\beta (E_{\vec{k}\uparrow}+E_{\vec{k}\downarrow})}{2}+2\cosh\frac{\beta (-E_{\vec{k}\uparrow}+E_{\vec{k}\downarrow})}{2}\Bigg),\nonumber
\end{eqnarray}
and also the free energy: $F/N=\Omega /N +\mu n$.

\section{Results}
We have performed an analysis of the influence of magnetic field on superfluidity, based on the equations (\ref{del} - \ref{pot}), paying special attention to the behaviour of the order parameter and the spin-up and spin-down electron density. In the following we set $t^{\uparrow}=t^{\downarrow}=t$ and use $t$ as the unit.

\begin{figure}[h!]
\begin{center}
\includegraphics[width=0.4\textwidth,angle=270]{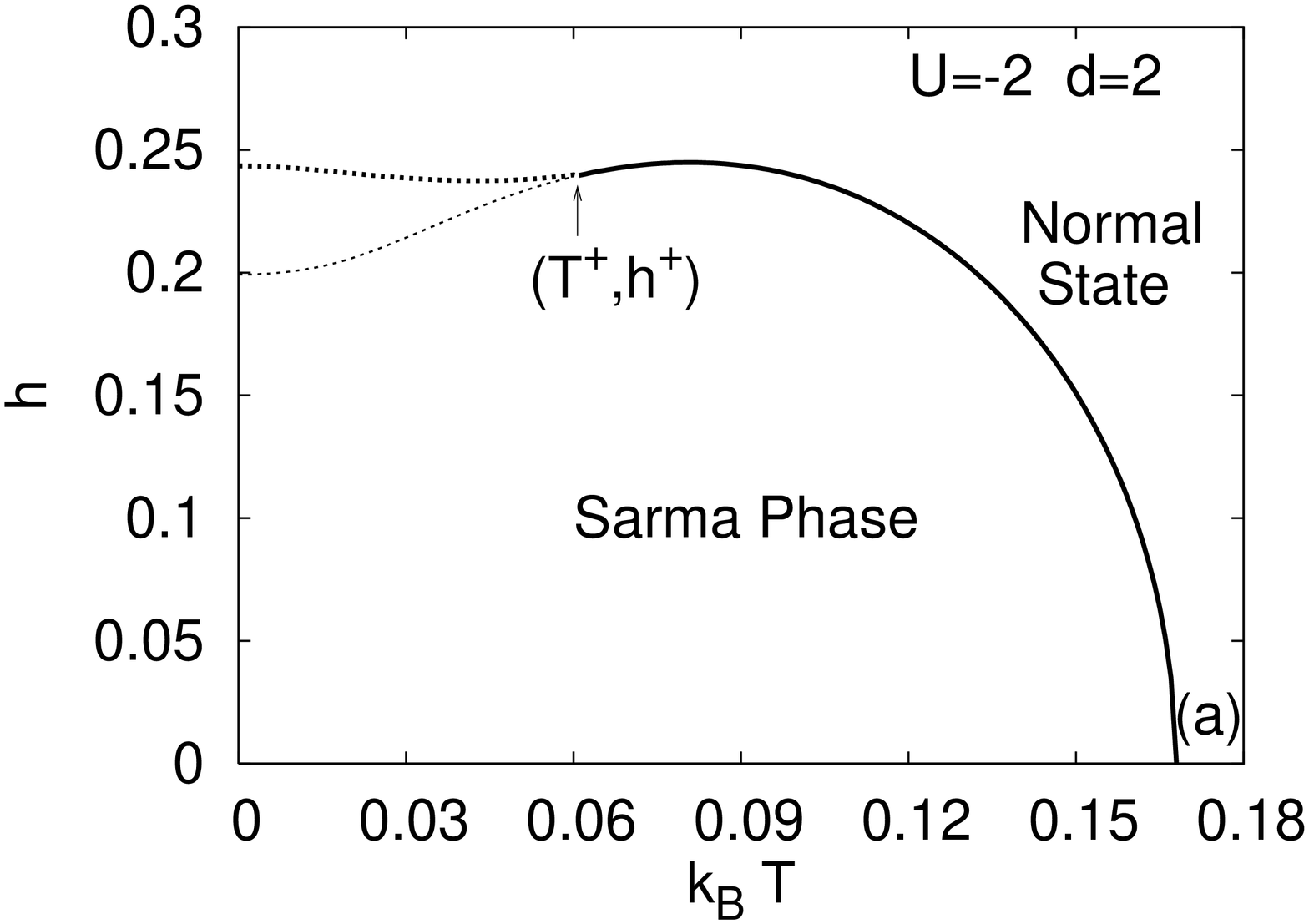}
\includegraphics[width=0.4\textwidth,angle=270]{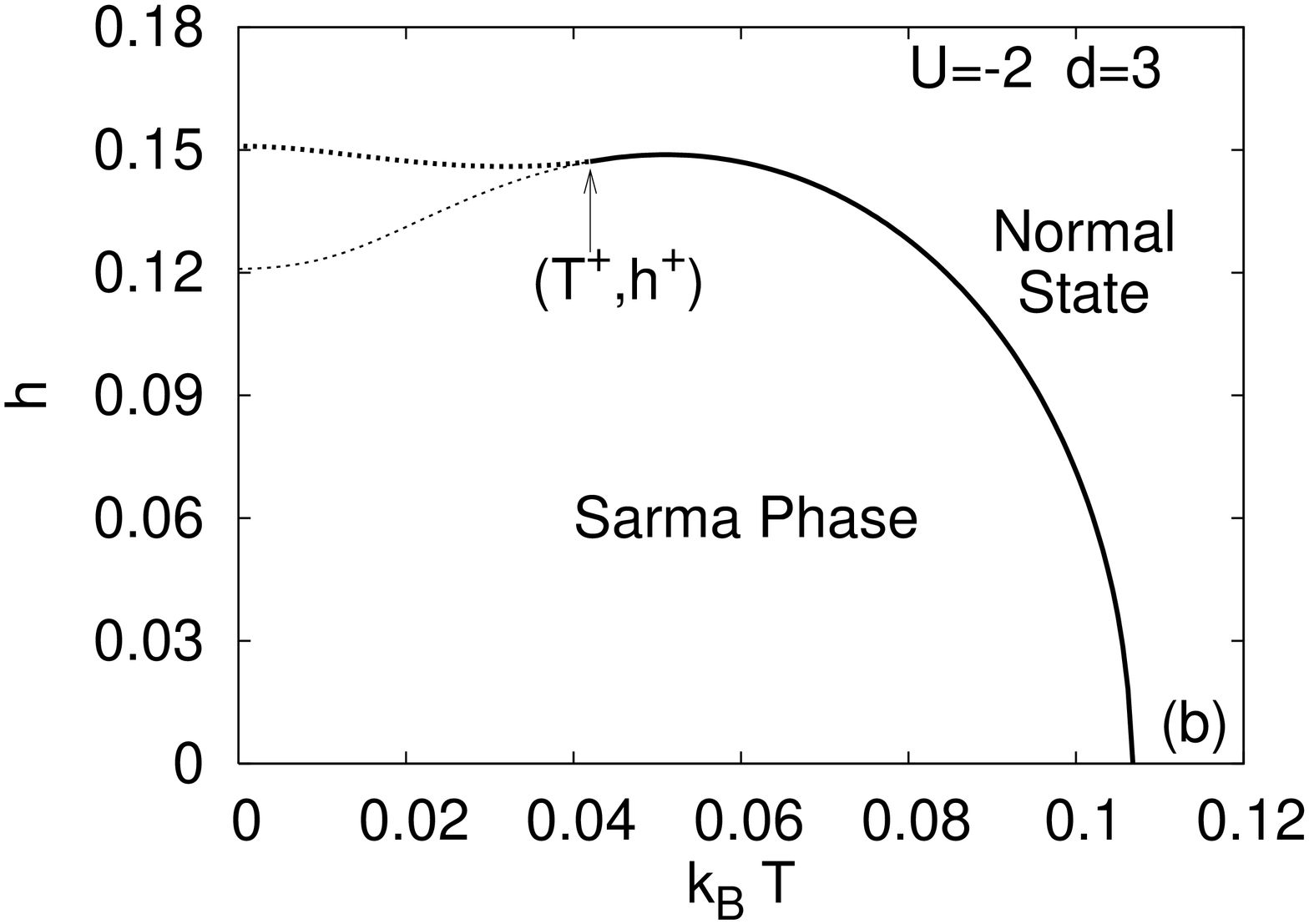}
\caption{Temperature vs. magnetic field phase diagrams for $U=-2$. (a) 2D lattice, $\mu =-1.358405$, (b) 3D lattice, $\mu =-1.627404$. The arrows indicate the tricritical point; the thick dotted line denotes the first order phase transition to the normal state. The chemical potentials have been chosen to yield $n \approx0.75$ at $T=0$ and $h=0$.}
\end{center}
\end{figure}

Fig. 1 shows the temperature vs. magnetic field phase ($T-h$) diagrams for the 2D (a) and the 3D (b) lattices for a fixed chemical potential $\mu$. Both in 2D and in 3D, a finite temperature second order phase transition takes place to the normal state at sufficiently low values of the magnetic field. With increasing magnetic field, the character of the transition between the superconducting and the normal state changes from the second order to the first order (the thick dotted line), which starts from the tricritical point (TCP). The line of the first order transition has been determined numerically from the condition: $\Omega_s =\Omega_n$ (where $\Omega_n$ and $\Omega_s$ denote the grand canonical potential of the normal and the superconducting state, respectively.). 
The curve below the first order transition line on the phase diagrams (the thin dotted line) is merely the extension of the line of the second order transition below TCP. The critical values of the magnetic field for a simple cubic lattice are lower than for the square lattice. The Hartree term raises the values of the critical magnetic field at $T=0$ for the first order transition, which exceed the Clogston limit \cite{C} ($h_{c}=\Delta_{0} \slash \sqrt{2}$, where $\Delta_{0}$ is the gap at $T=0$, $h=0$). 

It is important that the above diagrams were obtained for a fixed chemical potential. If the number of particles is fixed instead and $n\neq 1$, one will obtain two critical magnetic fields on the $T-h$ phase diagram \cite{A}, \cite{daniel}. The two critical fields define the phase separation region between the superfluid phase with the particle density $n_s$ and the normal state with the density of particles $n_n$. We note that, for a simple cubic lattice and for a weak attraction (Fig. 1b, $U=-2$), the differences between the values of these two critical fields are negligible and in this case practically only one critical field exists. 

\begin{figure}[h!]
\begin{center}
\includegraphics[width=0.4\textwidth,angle=270]{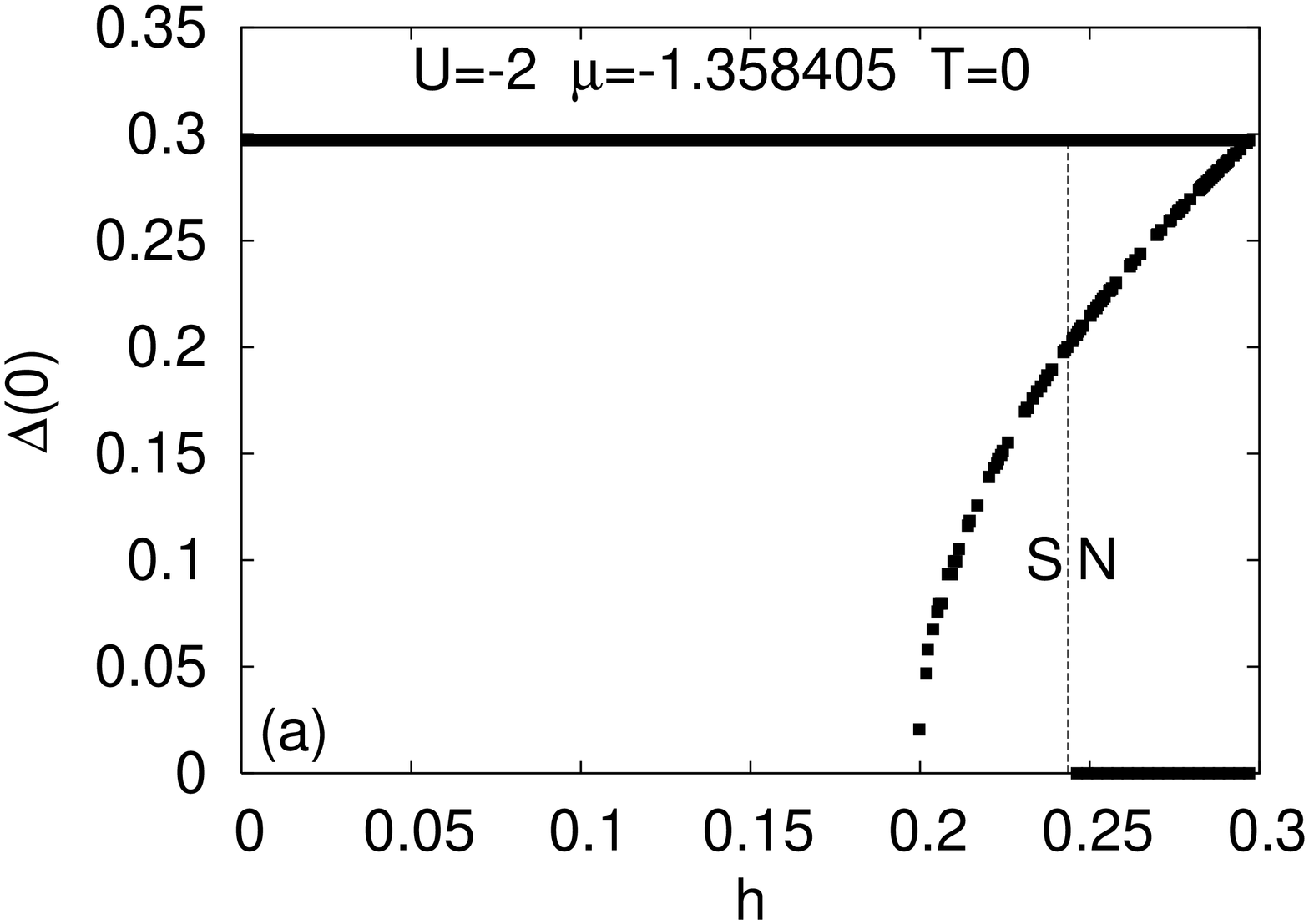}
\includegraphics[width=0.4\textwidth,angle=270]{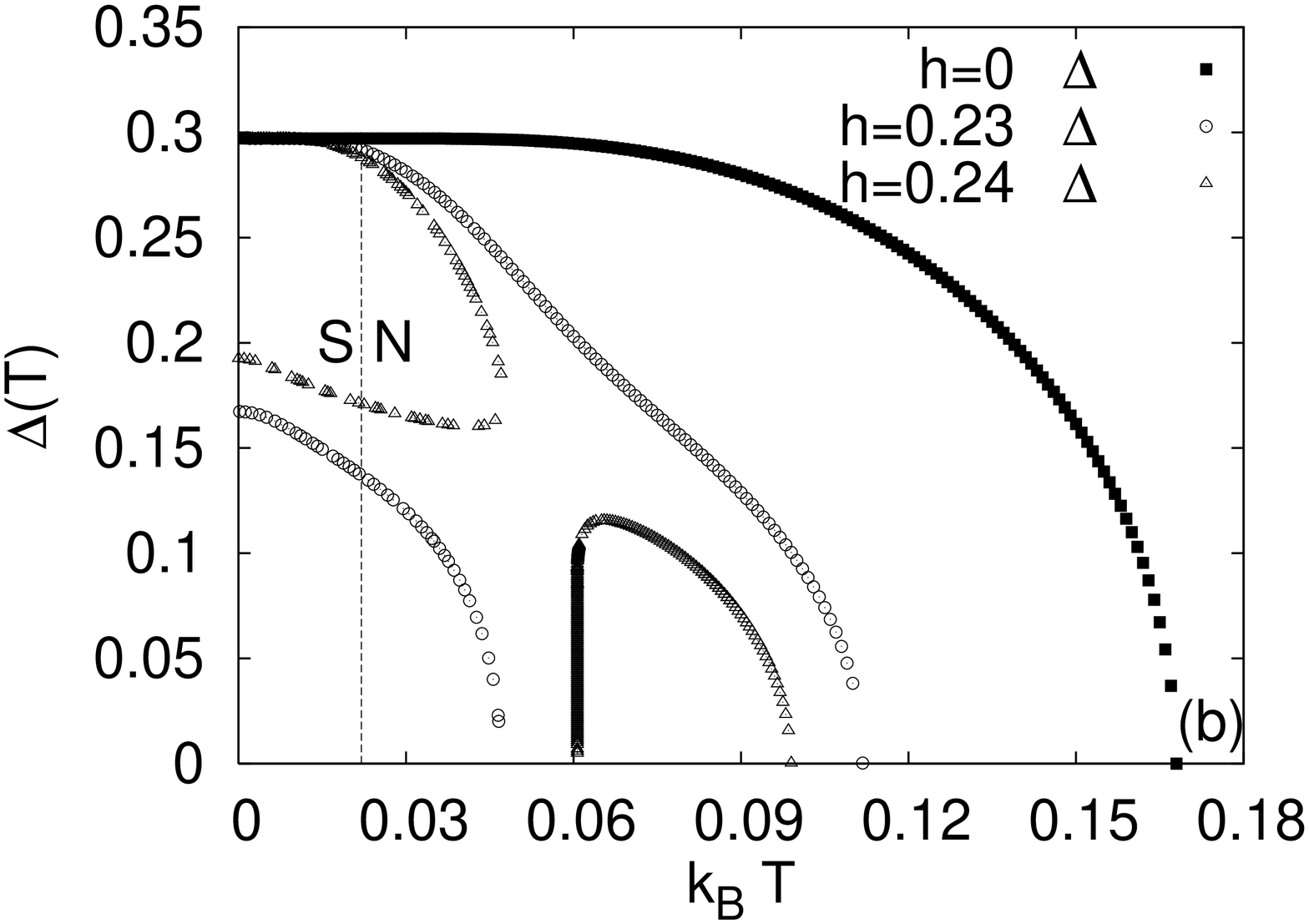}
\caption{Dependence of the order parameter on magnetic field at $T=0$ (a) and temperature (b) (for three fixed values of the magnetic field), for 2D and $U=-2$, $\mu=-1.358405$. In Fig 2b, for $h=0.23$ the lower branch is unstable. For $h=0.24$ the vertical dashed line denotes the first order phase transition to the normal state and there are two second order transitions in the reentrant case (at $T=0.06067$ and $T=0.09891$).
}
\end{center}
\end{figure}

\begin{figure}[h!]
\begin{center}
\includegraphics[width=0.4\textwidth,angle=270]{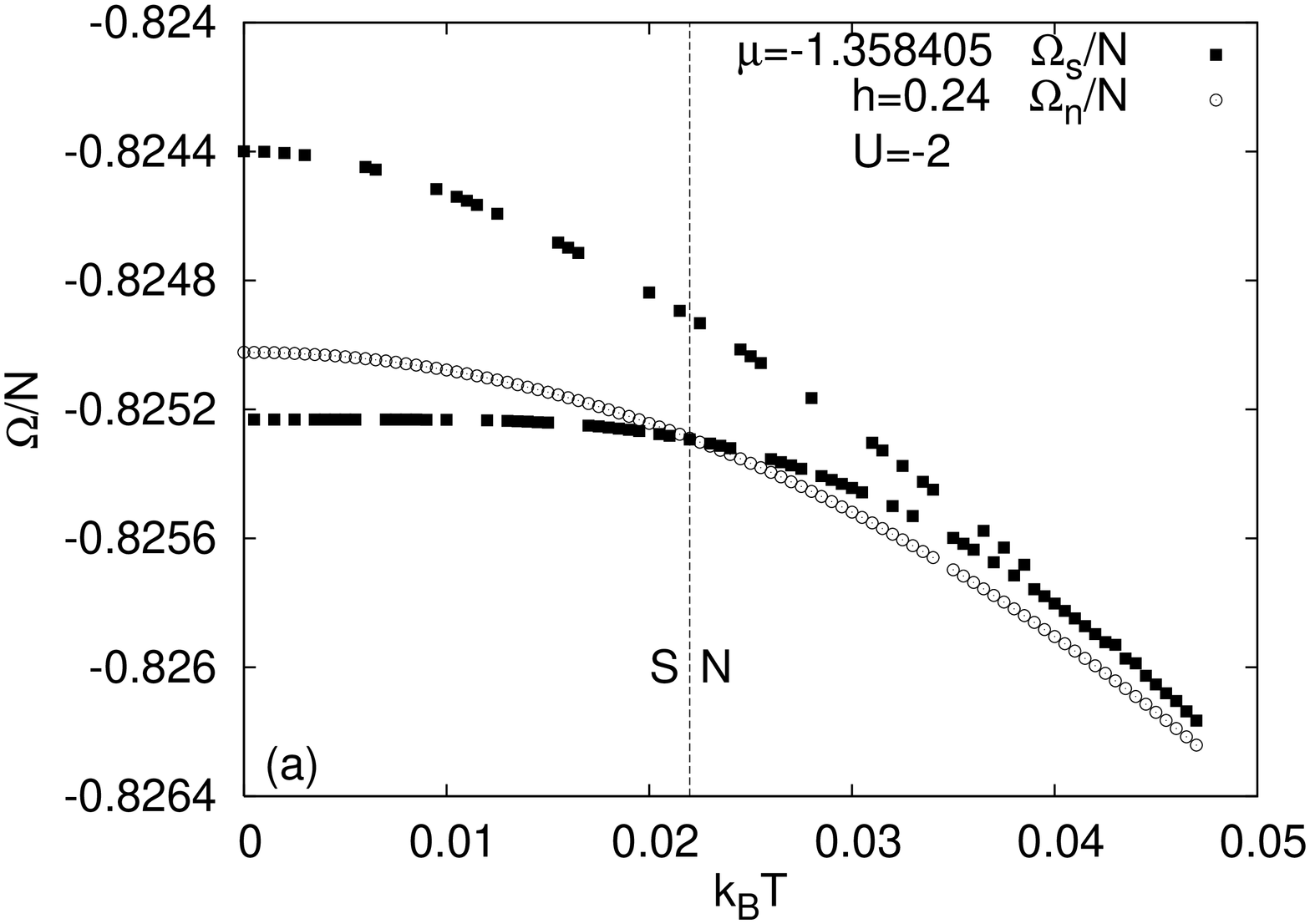}
\includegraphics[width=0.4\textwidth,angle=270]{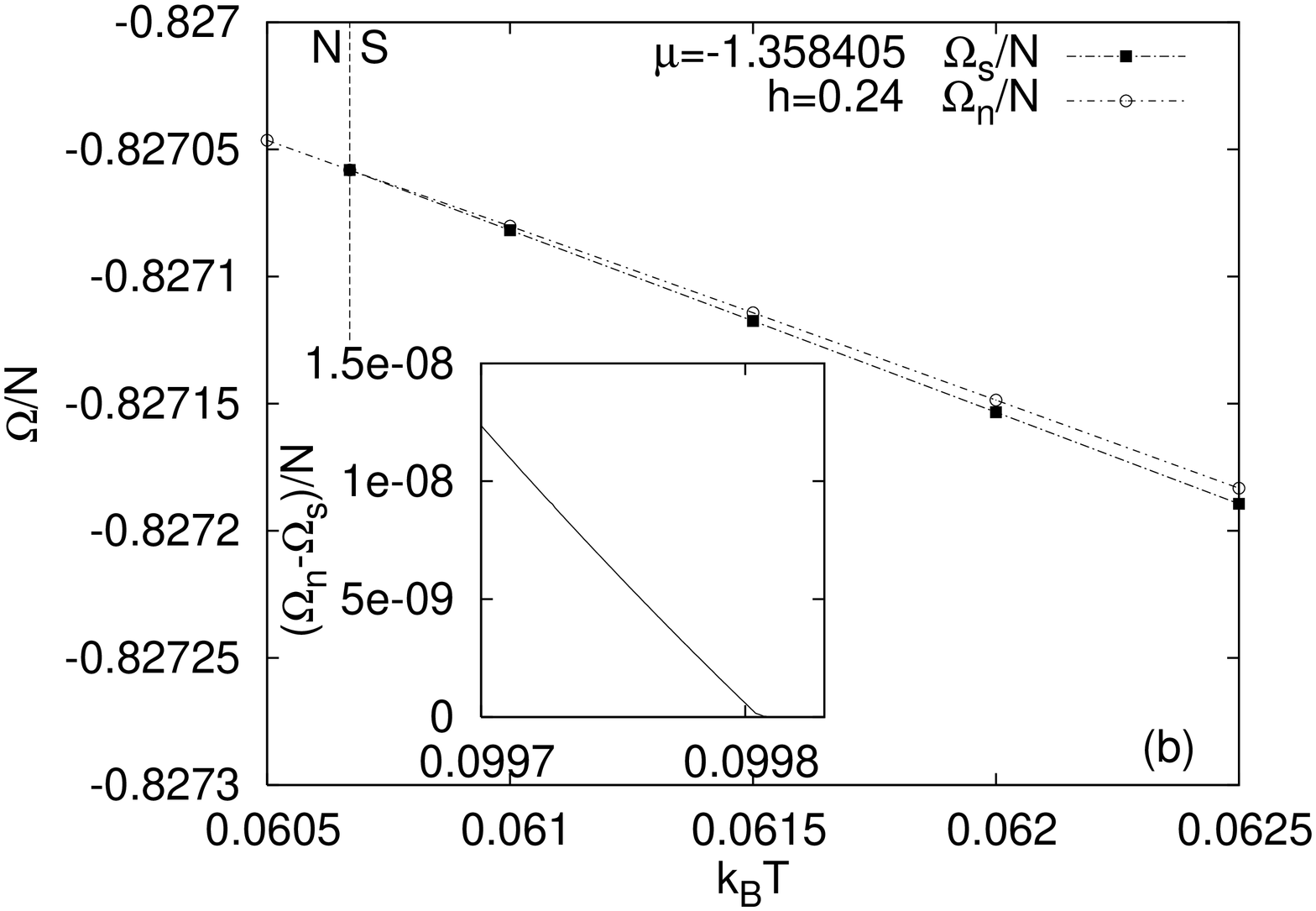}
\caption{Grand canonical potential vs. temperature for $h=0.24$ (2D case), (a) the first order phase transition to the normal state, (b) the details of the reentrant transition: second order transition from NS to SC and from SC to NS (inset). The vertical dashed lines mark the phase transition temperatures.   
}
\end{center}
\end{figure}

The change in the character of the transition is clearly visible in the behavior of the order parameter as a function of temperature, which is depicted in Fig. 2b. We have the first order BCS to NS transition at $T=0$ (Fig. 2a). For low magnetic fields, $\Delta$ vanishes continuously with increasing temperature. There arise two non-zero solutions for the order parameter, for $h=0.23$. The lower branch is unstable. 
We have also found a very interesting behavior of the order parameter for fixed $h=0.24$. In this case $\Delta$ vanishes discontinuously at $T=0.0225$ (the solutions with $\Delta \neq 0$ become energetically unfavorable, i.e. the upper branch is metastable and the lower branch is unstable for $T>T_c$.) -- thus we have the first order transition. However, for $T\geq 0.06067$ the superconducting solution becomes energetically favorable again (second order transition to the SC state) and $\Delta$ vanishes continuously (second order phase transition to the NS at $T=0.0998$), which is clearly visible in Fig. 3b.  Such behavior points out that for sufficiently high fields a reentrant transition takes place. Hence, the increase in temperature can induce superconductivity. Similar results have been obtained for the 3D case (and $U=-3$). The behavior of the grand canonical potential (Fig. 3a) indicates the point of the first order transition from the superconducting (Sarma Phase (SP)) to the normal state. 

\begin{figure}[h!]
\begin{center}
\includegraphics[width=0.4\textwidth,angle=270]{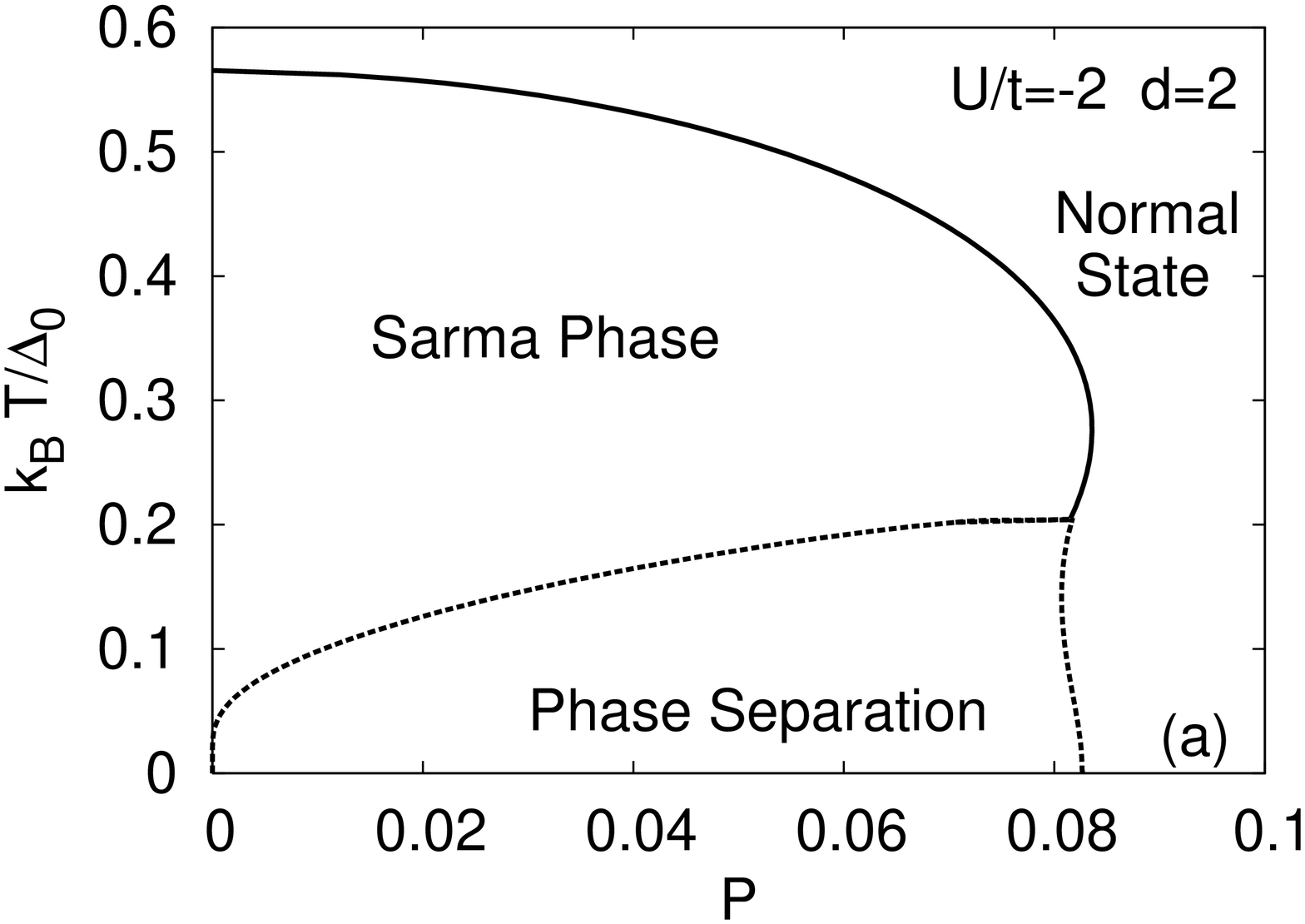}
\includegraphics[width=0.4\textwidth,angle=270]{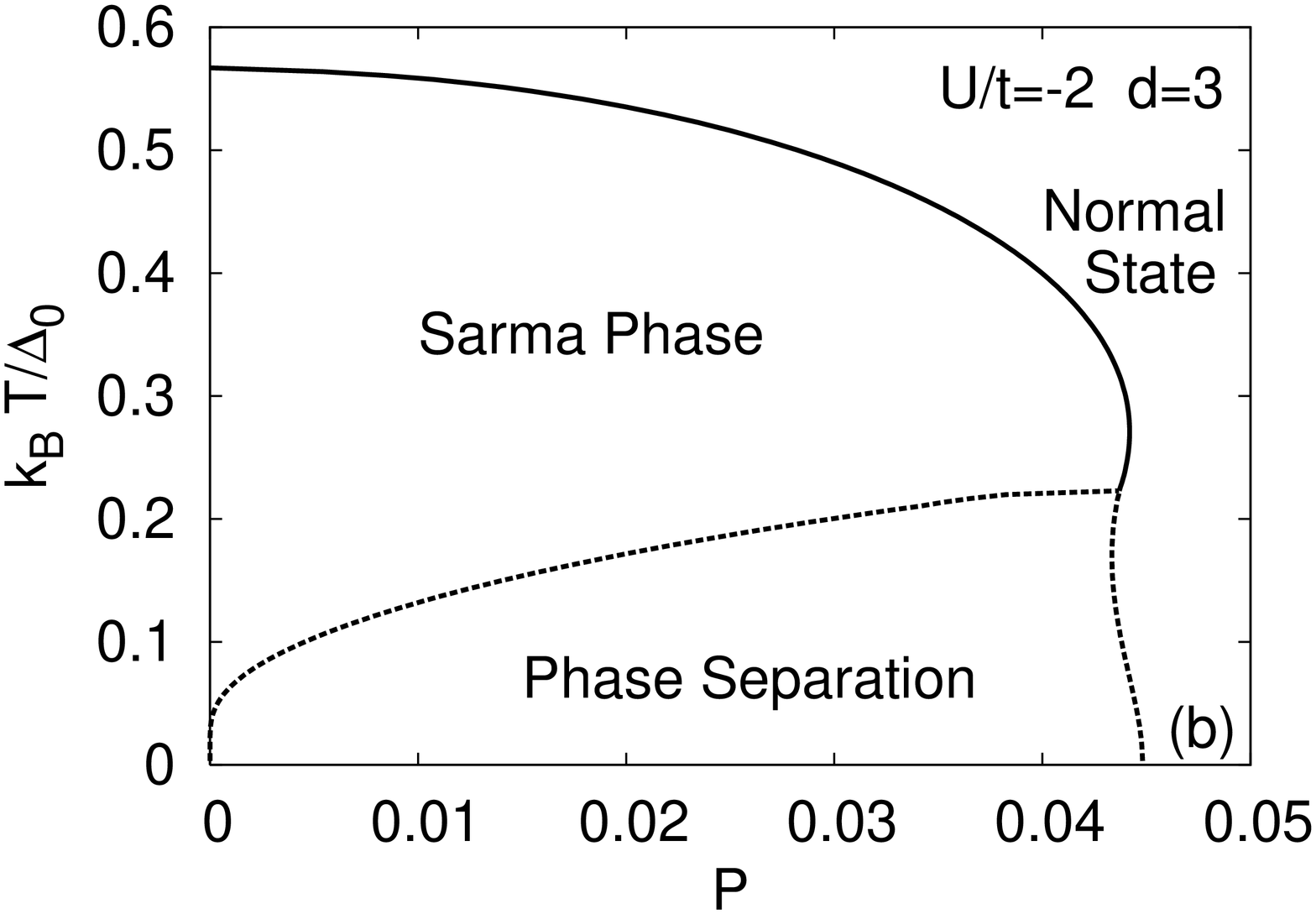}
\caption{Temperature vs polarization phase diagrams for $U=-2$. (a) 2D lattice, $\mu =-1.358405$, (b) 3D lattice, $\mu =-1.627404$. The dotted lines are the first order phase transition lines, $\Delta_{0}$ denotes the gap at $T=0$ and $P=0$. In (b) $n\approx 0.75$.
}
\end{center}
\end{figure}

Figure 4 shows the temperature vs. spin polarization ($T-P$) phase diagrams for the 2D case (a) and the 3D case (b). ($P=(n_{\uparrow}-n_{\downarrow}) \slash (n_{\uparrow}+n_{\downarrow})$). One can distinguish three states for the system under consideration: SP, Phase Separation region (PS) and NS. PS terminates at TCP. At $T\geq 0$ and $P=0$ we have the BCS phase. At $T\neq 0$, $\Delta \neq 0$ and $P\neq 0$, the system is in the Sarma phase (i.e. superconductivity in the presence of the spin polarization). The transition from SP to NS is of second order. Furthermore, we have a transition to the PS region (at $T=0$ the coexistence of the superconducting (BCS state, $\Delta \neq 0$, $P=0$) and the normal state ($\Delta =0$, $P\neq 0$), and for $T>0$ the coexistence of SP ($\Delta \neq 0$, $P\neq 0$) and paramagnetic NS ($\Delta=0$, $P\neq0$)). Because $\Delta$ vanishes discontinuously for higher magnetic fields (larger $P$), the transitions both from SP to the PS region and from the PS region to the NS are of first order (the dotted lines). Moreover, in the PS region not only the polarizations but also the particle densities in the SC and NS are different.

\section{Conclusions}
The influence of the magnetic field on superfluid properties of the attractive Hubbard model in a weak coupling regime has been considered. The $T-h$ and $T-P$ phase diagrams have been obtained for the two- and three- dimensional cases. We can distinguish following phases on the diagrams: the Sarma Phase, the Phase Separation region (between the spin-polarized superconducting state (with $n_s$) and the normal state (with $n_n$)) and the Normal State. In the presence of the magnetic field the densities of states are different for the particles with spin down and spin up. Moreover, the magnetic field destroys the superfluidity through the paramagnetic effect (or by a population imbalance).
One can also have a reentrant transition for sufficiently high magnetic fields on the $T-h$ and $T-P$ phase diagrams. For 2D system at $h=0$, the transition from the superconducting to the normal state is of the Kosterlitz-Thouless (KT) type. In the case under analysis, the phase fluctuations have been neglected. However, for weak attraction the Hartree-Fock critical temperatures may be comparable with the KT ones.  

In this paper, we have restricted the analysis to the case of $s$-wave pairing only with $\vec{q}=0$, leaving out the case of the FFLO phase. An important limitation, as far as the charged superfluid is concerned, is the neglect of the orbital effect of a magnetic field and the magnetic field fluctuations \cite{czart}. 
Investigation of the competition between the superconducting phases and CDW diagonal ordering (in particular at and close to the half-filled band $n=1$) would also be an interesting extension of this paper.

\section*{Acknowledgements}
We would like to thank Prof. S. Robaszkiewicz for valuable discussions.

\end{document}